\documentstyle[11pt,ska,twoside,psfig]{article}
\begin{document}
\title{SETI with SKA}
\author{Alan J. Penny}
\affil{ Rutherford Appleton Laboratory, Chilton, Didcot OX11 0QX, 
\newline United Kingdom}


\begin{abstract}

SETI with SKA would be by far the most powerful SETI search ever
undertaken, covering enough stars with enough sensitivity to probe
significantly further towards those other Earth civilisations than
previous ones. This paper discusses the rationale behind radio SETI
searches, and explains why SKA will be such as big step forward and
will make few demands on telescope time.

\end{abstract}

\section{Introduction}

This paper discusses the use of SKA in the search for
extra-terrestrial intelligence (SETI). It assumes a SKA with a
collecting area of a square kilometre, and an angular resolution of
the individual antennae of about 3 degrees, with the synthesised beam
having arcmin resolution. It is thus not concerned with the extended
baseline components. The spectral range is assumed to cover 2 cm to 1
metre, with the long wavelength limit determined by the value of
covering wavelengths used by powerful terrestrial transmitters. The
need for a radio-quiet location is strong.

\section{What would SETI with SKA look for?}

SETI looks for extra-terrestrial intelligence, but what sort of
intelligence?

Intelligence can be defined in many ways and covers wide ranges of
life. Intelligence defined as life capable of `problem solving' would
include bacteria and dinosaurs, but neither are the sort of SETI
intelligence we mean. Intelligence as life with `tool making and
using' would include crows, apes and Neanderthal man, but again these
do not seem to be suitable as SETI targets. Intelligence as life with
a `technology-based civilisation' limits us to the appearance of
Cro-Magnon man, and even here we have to be careful. The evidence for
Cro-Magnon civilisation before about 7,000BC mainly rests on cave
paintings, and it has been pointed out (Humphrey, 1998) that these
bear many similarities to the art of some autistic children and thus
could indicate a `civilisation' lacking in many of the qualities of
what we would consider as a `civilisation'. It might seem that we
would be safe after 7,000BC, with the civilisations of the Near-East,
but even here it has been controversially postulated (Jaynes, 1990)
that it was not until about 1000BC that clear evidence for
civilisations like our own can be found. An example of the reasons
adduced for such a limitation is that while the technology-based
civilisation of Ancient Egypt certainly knew about the 3,4,5 rule for
right-angled triangles, there is no evidence for any comprehension of
a science behind this geometry. It seems that we can only be safe by
saying `life with the characteristics of Homo Sapiens civilisations
over the last 3000 years'.

Fortunately, for SKA we can be even more restrictive. SKA would only
be able to find civilisations that emit radio waves which are
distinguishable from those from natural sources. This is not to say
that other types of civilisations are not of interest, just that
radio telescopes could not find them. So for SKA we define SETI as
the search for `civilisations that emit radio waves which are
distinguishable from those from natural sources'.

Present `radio SETI' searches discriminate between natural and ETI
emission by looking for very narrow (about 1Hz) band emission. Not
only is such radiation unlikely to be naturally produced, but we know
that our own civilisation radiates such signals. TV stations have a
narrow-band carrier wave of this frequency narrowness, with the
actual signal spread over a MHz about that. About 1/3rd of the power,
or typically 200kW, is in this carrier wave, at frequencies of the
order of 0.5GHz. Thus we at least know that civilisations can emit
such signals, strong enough to be detectable.

\section{Is this type of radio SETI the best SETI search method?}

Would it be better to look for optical emissions? For the thermal
signatures of Dyson spheres? Should we look for coded radio signals?
The question can be sorted into two cases, civilisations like our own
and civilisations unlike our own. For the first case, narrow-band
radio emission is indeed the strongest artificial signal from us, and
so this type of radio SETI is best. Further, we can at least make
some discussion of the possible number of such civilisations. For the
second case, as we cannot even make a sensible discussion of the
likelihood of different types of civilisations and their emissions,
then no type of SETI search can be classed as `better' than any
other. Radio SETI would be just as good as any other type.

Thus, overall, radio SETI is the best method, because it alone looks
for both known and unknown types of civilisations. However, there is
certainly a case that other types of searches would be of value, as
they will seek other classes of unknown civilisations, on the basis
of `if you don't look, you won't find'.

\section{What are the chances of a radio SETI-type civilisation existing?}

Civilisations that emit radio waves could conceivably come in many
forms, such as combinations of whirling mini-Black Holes or
structures in interstellar clouds. However, as described above, we
have no understanding of such forms, and so can make no sensible
discussion on the probability of their occurrence.

We do know something about our own intelligence, so can we deduce
anything about the chances of similar civilisations appearing
elsewhere? Unfortunately, it seems that even here only a rough
discussion can be done. We don't understand how probable it is that
life starts on an Earth, or how, having started, what the probability
is that it leads to a civilisation.  [This omits the deeper problem
that we don't even understand what life and intelligence `are'. For
an exploration of this problem, see e.g. Cairns-Smith (1998).] There
are many opposing points of view on these probabilities (see e.g.
Ward and Brownlee, 2003a and 2003b).

First of all, we have to have a planet like the Earth with life on
it. We do not yet know how common such Earths are. Recent searches
have found gas giant planets orbiting other stars, but are not
powerful enough to detect terrestrial planets, and the formation
theories (e.g. Wetherill, 1996) for such planets contain many
unknowns. But the Eddington 
\footnote{http://sci.esa.int/home/eddington/index.cfm} and Kepler
\footnote{http://www.kepler.arc.nasa.gov} space telescopes in this decade will
determine the frequency of occurrence of terrestrial planets. Having
got an Earth, what are the chances of it developing life? We do not
understand how life starts, but much information on the frequency of
simple life will come in the next decade from the Mars missions and
from the Darwin/TPF \footnote{http://sci.esa.int/home/darwin/index.cfm}
\footnote{http://planetquest.jpl.nasa.gov/Navigator/tpf\_nav.cfm} space
telescope for Earths orbiting other stars.

Having got an Earth with life, what are the chances of a
civilisation? Here we have neither a theoretical understanding nor a
non-SETI way of making an estimate. We know (see e.g. Ward and
Brownlee, 2003a) that life started early on our Earth, within about
0.5 Gyr after its birth, and that our civilisation has occurred close
to the end of the period when life can exist on the planet's surface
(see e.g. Ward and Brownlee, 2003b). These timescales are compatible
with a number of cases. It would fit for simple life being common,
but that the average timescale for civilisations is comparable with
the lifetime of stars and thus that (say) from 10\% to 90\% of
suitable planets have a civilisation phase. It is, statistically
speaking, more compatible with simple life being common, but
civilisations taking a very long time to arise, and thus being very
rare. It is even compatible with both simple life and civilisations
being rare. The most uncertain part of this is the problem that we
only know about life and civilisations here on Earth because we are
around to observe them. This `Anthropic Principle' complication is
well covered in Bertola and Curi (1993).

Thus we cannot at present reliably estimate the probability that a
star will have an Earth-like planet where life will reach a
civilisation phase.

Moreover, we do not know how long such a civilisation phase would
last on average. We know that we have been emitting radio waves for
about 100 years. How long will this phase last? Clearly we cannot
estimate this, but a hand-waving argument might go that judging by
the changes in technologies in the past we will move onto another
phase no later than (say) 1000 years in the future. It seems unlikely
that we have now reached the ultimate technology in communications.
Such a brief timescale means that, as stars live for billions of
years, even if each has a planet that reaches a civilisation phase,
only one in a million stars will have a planet in that phase at any
one time.

There is a further complication -- the Fermi Paradox. If our
civilisation is not the only one in our Galaxy, then it seems
inescapable that advanced civilisations will have developed the
ability, and have had enough time, to come here or make their
presence known, very soon after the start of their `radio
wave-emitting civilisation' phase. As not one seems to have done so,
then either we are alone, or there is some `Cosmic Zoo' that prevents
the existence of such civilisations from being known by us. This Zoo
might even prevent us from seeing `unadvanced' civilisations in our
own phase. Many possible reasons and mechanisms for such a zoo have
been proposed, some of which would rule out any SETI search
succeeding, and some of which only explain the previous null results
from SETI searches. However, none of these solutions seem to be
inescapably valid. So we have to leave the paradox unresolved, with a
mental note that, on the face of it, it points to us being alone.

\section{Other radio SETI searches}

Radio SETI searches first started in 1960s, but the most powerful
searches have been the BETA all-sky survey, and the Phoenix targeted
search. BETA (Leigh, 1998) used a 26-metre dish, covering 1.42-1.72
GHz with a resolution of 0.5Hz. It showed that in the northern half
of the sky there was no narrow-band source, with the brightness of a
typical Earth TV station, within 0.05 pc. The ongoing Phoenix search
(Backus et al, 2001), using a variety of telescopes from 50 to
300-metres, has looked at 1000 stars covering a band from 1 to 3 GHz,
and again found no source such as bright as an Earth TV station at
0.5pc (although of course the stars surveyed were all more distant
than that).

The privately funded Allen Telescope Array (ATA) (Tarter, J. et al,
2002), operational in 2005, with a collecting area of one hectare,
will cover 100,000 stars over a 0.5GHz band, with a sensitivity equal
to the Phoenix search

\section{SETI with SKA}

SKA, like ATA, will use the fact that, in the 3-degree FOV of the
single elements, up to 10 1-arcmin beams can be synthesised and 10
stars observed. This can be going on, `piggy-back', while the
astronomy target is observed. Thus apart from the cost of the
synthesiser, with its multiple target, wide band, narrow bandwidth,
coverage, the SETI search comes for free. Of course, not all the sky
will be covered with the astronomy pointings, so some dedicated
fraction of observing time may also be needed to cover nearby, or
specially interesting, stars.

The default program is to have receivers with 0.01Hz frequency bins
covering the 0.5-10 GHz band. This band is the quiet spectral region
between the non-thermal galactic background and the atmospheric
emission bands. With 100 seconds on each target (if the astronomy
observation lasts longer, then new beams are made and new stars
looked at), and 3 re-looks per star (to deal with false alarms), then
one million stars can be observed within 10 years, with a sensitivity
to see that Earth TV station at 3 pc.

The main gains with SKA are the ability to cover one million stars,
with an increased sensitivity. As discussed above, this means that
for the first time we would have some chance of catching a
civilisation in its presumably brief radio-emitting phase. This
chance will depend on the extent to which stars have habitable
Earths, the extent to which each develop a civilisation, and the
extent to which civilisations have radio emitters significantly more
powerful than present day Earth ones (as practically all the SKA SETI
stars will be significantly further away than 3 pc).

SETI with SKA will be by far the most powerful SETI search ever
undertaken, covering enough stars with enough sensitivity to probe
significantly further towards those `Earth'-type civilisations. And
also do a wider and deeper for search for those `other'-type
civilisations.

\acknowledgements

I am grateful for information and advice from Jill Tarter and Ian Morrison.

\clearpage
.

\end{document}